\documentclass[prb,reprint,superscriptaddress]{revtex4-1}

\usepackage{amsmath}
\usepackage{amssymb}

\usepackage{graphicx}
\graphicspath{./}

\usepackage{hyperref}
\hypersetup{
 bookmarks=true,		% show bookmarks bar?
 unicode=false,			% nonLatin characters in Acrobat’s bookmarks
 pdftoolbar=true,		% show Acrobat’s toolbar?
 pdffitwindow=false,		% window fit to page when opened
 pdfstartview={FitH},		% fits the width of the page to the window
 pdfcreator={pdflatex},		% creator of the document
 pdfproducer={vim},		% producer of the document
 pdfkeywords={quantum chemistry} {chemical bond} {electron-deficit bonds} {quantum correlations} {entanglement} {multiorbital correlations} {diborane} {C-Be-C complex}, % list of keywords
 pdfnewwindow=true,		% links in new window
 colorlinks=true,		% false: boxed links; true: colored links
 linktoc=page,			% defines which part of an entry in the table of contents is made into a link
 linkcolor=blue,		% color of internal links	(red)
 citecolor=blue,		% color of links to bibliography
 filecolor=blue,		% color of file links
 urlcolor=blue			% color of external links
}

\newcommand{\m}{\scalebox{0.75}[1.0]{$-$}}

\begin{document}
\title{Sensitivity of Coupled Cluster Electronic Properties on the Reference Determinant: Can Kohn-Sham Orbitals Be More Beneficial than Hartree-Fock Orbitals?}
\date{\today}
% **********************
\author{Zsolt Benedek}
\affiliation{
Wigner Research Centre for Physics, H-1525, Budapest, Hungary}
\affiliation{
Department of Inorganic and Analytical Chemistry, Budapest University of Technology and Economics, H-1111 Budapest, Hungary}
\affiliation{Department of Chemical and Biological Engineering, The University of Alabama, Tuscaloosa, AL-35487, USA}
% **********************
\author{Paula T\'im\'ar}
\affiliation{
Wigner Research Centre for Physics, H-1525, Budapest, Hungary}
\affiliation{
Department of Inorganic and Analytical Chemistry, Budapest University of Technology and Economics, H-1111 Budapest, Hungary}
% **********************
\author{Tibor Szilv\'asi}
\affiliation{Department of Chemical and Biological Engineering, The University of Alabama, Tuscaloosa, AL-35487, USA}
% **********************
\author{Gergely Barcza}
\email{barcza.gergely@wigner.mta.hu}
\affiliation{
Wigner Research Centre for Physics, H-1525, Budapest, Hungary}
\affiliation{Department of Chemical and Biological Engineering, The University of Alabama, Tuscaloosa, AL-35487, USA}
\affiliation{Department of Physics of Complex Systems, ELTE E\"otv\"os Lor\'and University, H-1117, Budapest,
Hungary}
% **********************

\begin{abstract}
Coupled cluster calculations are traditionally performed over Hartree-Fock reference orbitals (HF-CC methodology). 
However, it has been repeatedly argued in the literature that the use of a Kohn-Sham reference (KS-CC methodology) might result in improved performance relative to HF-CC at the same computational cost.
In the present theoretical study, we re-examine the relation of HF-CC and KS-CC methods by comparing the results of widely applied truncated CC calculations (CCSD, CCSD(T), CCSDT) to the limit of full configuration interaction (FCI), which – in contrast to wave-function diagnostics with vague physical meaning or experimental data with considerable uncertainty – serves as an undebatable reference point of accuracy. 
We select seven different sets of reference orbitals (HF and six different KS density functionals: LDA, BP86, M06-L, B3LYP, HSE06 and – for completeness - the Hartree formula) and four different basis sets (STO-3G, 6-31G, cc-pVDZ, cc-pVTZ) to perform the latter calculations on nine small molecules with increasing multi-reference character (BH, HF,  HF, CF, BF, NO, OF, BN, C$_2$, B$_2$). 
We find that apart from incidental exceptions, the Kohn-Sham referenced CC methods show systematic deterioration compared to HF-CC, that is, the KS-CC molecular properties (electronic energy and density) are always farther from the FCI limit than those obtained from HF-CC at the same coupled cluster level. 
Furthermore, the introduction of common approximations (frozen core, density fitting) to the CC calculation results in significantly higher errors in the case of KS reference. 
Additional analysis of high-level CC calculations on transition metal complexes suggest that these observations can be generalized to larger molecules. 
We conclude that the use of KS reference orbitals is not expected to increase the reliability of low-level CC  energetics.
Nevertheless, molecular errors from the components of the studied chemical reaction might fortunately cancel out resulting in illusory improvement compared to HF-CC. 
It is also notable that  the choice of reference orbitals has negligible influence on the results at sufficiently high CC levels which can be estimated by test calculations or by  the magnitude of double amplitudes.
Therefore, the application of KS-CC is not unreasonable as it might bypass the difficulties of HF convergence. 

\end{abstract}

\maketitle

\section{Introduction} 

Coupled cluster (CC) calculations~\cite{cizek1966} are considered to be one of the most reliable computational methods in quantum chemistry and thus often used as benchmarks~\cite{bartlett2007}. Their application, however, is strictly limited to single-reference systems, where the CC ansatz of
$    \Psi_{\rm CC} = e^T  \Phi_{\rm HF} $
contains a sufficient number of electron configurations to accurately describe the electronic structure, even if the excitation operator $T$ is truncated. Upon the increase of the multi-reference character of the system, that is, upon the decrease of the weight of the Hartree-Fock (HF) orbital~\cite{hartree1935} derived Slater determinant $\Phi_{\rm HF}$ in the ansatz, the reliability of the commonly applied truncated CC methods - such as CCSD, CCSDT or the gold standard CCSD(T)~\cite{raghavachari1989} - becomes increasingly questionable. From a practical point of view, this is a serious problem because several important homogeneous and enzyme catalysts relevant for various applications contain transition metal centers with strong multi-reference character. 

To tackle the latter issue, an alternative CC ansatz was proposed based on Kohn-Sham Density Functional Theory (KS-DFT).       
The KS-CC formalism, $\Psi_{\rm KS\m CC} = e^T \Phi_{\rm KS}$, differs practically from the original coupled cluster methodology in the reference state $\Phi_{\rm KS} $.
This Slater determinant  is constructed from KS-DFT derived auxiliary Kohn-Sham one-electron orbitals~\cite{tong1966} whose  physical interpretation is considered somewhat vague~\cite{stowasser1999}. 
Nevertheless, during the last two decades, a considerable number of KS-CC studies appeared in the literature \cite{harvey2003,beran2003,olah2009,chen2010,radon2011,vasiliu2015,fang2016,fang2017,feldt2019,bertels2021} which repeatedly argued that the use of KS reference determinants is not only permitted by the CC equations, but it might even lead to improved results due to the decrease of the multi-reference character in some aspects.

To the best of our knowledge, KS-CC calculations were proposed for the first time by Harvey and Aschi in 2003~\cite{harvey2003}. 
In their work on the interconversion of iron carbonyl complexes (Fe(CO)$_4$ + CO $\rightleftharpoons$ Fe(CO)$_5$, in various spin states), they initially utilized the usual HF-CCSD(T) methodology and calculated the $T_1$ diagnostic for the studied systems as a test of multi-reference character. 
This measure is defined as 
\begin{equation}
    T_1 = \frac{\| t_1 \|}{\sqrt{n_{\rm corr}}}\,,
    \label{t1}
\end{equation}
where $\| t_1 \|$ is the Euclidean norm of the vector of $t_1$ amplitudes, i.e., the coefficients of singly excited determinants in the CC ansatz, and $n_{\rm corr}$ denotes the number of correlated electrons~\cite{lee1989}. 
Empirically, $T_1 < 0.02$ is recommended for reliable CCSD(T) results but 0.04-0.06 was obtained for iron carbonyls. 
To gain more reliable energies at the same computational cost, the authors also performed CCSD(T) calculations using BP86 Kohn-Sham orbitals~\cite{perdew1986,becke1988}, which resulted in improved, but borderline (0.02-0.025) $T_1$ diagnostics. 
In addition, considerable deviations of up to 4 kcal/mol were observed between the respective HF-CCSD(T) and KS-CCSD(T) net reaction energies and the authors suggested the KS referenced results to be more accurate. 
Unfortunately, the latter hypothesis could not be tested against experimental energies which are only approximately determined.
Later, the same group also applied KS-CCSD(T) methodology for Heme model iron complexes, in the case of which HF-CCSD(T) calculations turned out to be practically unfeasible due to bad convergence behavior~\cite{olah2009} whereas the use of B3LYP Kohn-Sham orbitals~\cite{lee1988,stephens1994} eliminated the convergence issues. 
Furthermore, in the few cases where both KS and HF referenced results are available, it was observed that KS shows decreased singles and doubles CC amplitudes; the electronic energy, however, only negligibly ($< 1$ kcal/mol) altered compared to HF counterpart. 
In 2019, as their most recent publication related to the topic, the group extensively studied the effect of different settings in CCSD(T) calculations, including the choice of the reference orbital type~\cite{feldt2019}. 
They selected a C-H activation reaction on a non-heme type iron complex, i.e., [Fe(NH$_3$)$_5$O]$^{2+}$ + CH$_4$ $\rightarrow$ [Fe(NH$_3$)$^5$O-H-CH$_3$]$^{2+}$ in various spin states, as model system and found significant, up to 4 kcal/mol deviations between HF-CCSD(T) and KS-CCSD(T) energetics. 
Interestingly, the choice of the KS functional (B3LYP, BP86, TPSS~\cite{tao2003} and M06~\cite{zhao2006,zhao2008} were tested) appeared to be practically insignificant, as the different KS-CCSD(T) energies varied within a mere 1 kcal/mol. 
Again, the authors suggested the KS reference to be more probably reliable without external benchmarking.
In the meantime, relying on the above works, Shaik~\cite{chen2010} and Pierloot~\cite{radon2011} also published B3LYP referenced coupled cluster studies on various iron complexes. 
These papers already assumed that the KS-CCSD(T) approach is equally reliable to HF-CCSD(T), thus, the traditional HF referenced calculations were not even performed.

Moving beyond iron, the Dixon group investigated the compounds of numerous $d$ and $f$ transition metals using both HF-CCSD(T) and KS-CCSD(T). 
At first~\cite{vasiliu2015}, the group explored the energy profile of water addition to metal dioxides, i.e., MO$_2$ + H$_2$O $\rightarrow$ M(OH)$_2$O), and found that the change of reference orbitals (Hartree-Fock or PW91 Kohn-Sham~\cite{perdew1992}) results in deviations of up to 5 kcal/mol in the energy level of intermediates and transition states. 
Since the $T_1$ diagnostic was considerably smaller with KS reference, the ambiguity between the methods was interpreted as the improvement of CCSD(T) energetics by the inclusion of KS orbitals. 
Unfortunately, experimental data cannot be used for the assessment of accuracy in this case, as they are only available for metal oxide surfaces, rather than for separate dioxide molecules.
In their subsequent works~\cite{fang2016,fang2017}, novel computational protocols were developed for the prediction of the heat of formation of transition metal hydrides, oxides, sulfides and chlorides, which treats the electronic energy at CCSD(T) level of theory. 
Again, it was observed that the use of PW91 or B3LYP orbitals instead of the usual HF orbitals decreases the $T_1$ diagnostic and also alters the energy (or the derived enthalpy) by up to several kcal/mol. 
In this case, although experimental dissociation energies and enthalpies are abundantly available in the literature, most of these come with error bars exceeding the orbital choice related deviation. 
The Dixon group still managed to find one example where the B3LYP-KS reference appears to be unequivocally superior to HF, i.e., the enthalpy of the reaction of UCl$_6$ + 3F$_2$ $\rightarrow$ UF$_6$ + 3Cl$_2$, experimentally measured to be 278.0$\pm$1.2 kcal/mol at room temperature, was predicted to be 285.9 kcal/mol and 277.8 kcal/mol by HF-CCSD(T) and KS-CCSD(T) based protocols, respectively~\cite{fang2016}.
Even though KS-CCSD(T) clearly produces better numerical agreement,  assessing the quality of the computational models for the single benchmark is still challenging. 
In fact, not only an incidental cancellation of numerical errors cannot be ruled out, but  thermochemical correction factors  of the temperature-dependent enthalpy  might also be a source of error.

Altogether, in the above pioneering KS-CC papers, the practicality of the KS based approach is typically argued on the basis of favorable $T_1$ diagnostics. 
This is however a questionable argument as the usefulness of $T_1$ as a measure of the quality of CC energy is disputed~\cite{liakos2011}. 
In contrast, the Head-Gordon group, for example, compared the performance of HF-CC and KS-CC calculations based on the reproduction of experimental vibrational frequencies \cite{beran2003,bertels2021}. 
On a database of diatomic molecules, KS-CCSD(T) was found to outperform HF-CCSD(T) in accuracy by nearly a factor of 5 regardless of the density functional, among which BLYP~\cite{becke1988,lee1988}, B97M-rV~\cite{mardirossian2017}, B97~\cite{becke1997}, $\omega$B97X-V~\cite{mardirossian2014} and $\omega$B97M-V~\cite{mardirossian2016} were tested. 
This result was attributed to the smooth change of KS orbitals with nuclear displacement  contrasted to the rapid change of HF orbitals. 
Thus, the less efficient frequency calculations with HF-CCSD(T) do not necessarily indicate less accurate electronic energies at the optimal ground state nuclear configuration, which is of primary interest in typical coupled cluster studies.

In view of all these results, in order to  validate the suspected favorable performance of the truncated KS-CC approaches with unquestionable mathematical rigor high-level numerical reference quantities are needed.  
As the truncated CC methods approximate the full configuration interaction (FCI) wave function, the upper limit of their accuracy is the FCI energy for the applied basis set which is invariant to the chosen reference orbitals, see Appendix for details.  
Accordingly, the main goal of the present work is to compare the truncated KS-CC and HF-CC results to calculations of essentially FCI quality benchmark results, which - in contrast to the ambiguous experimental data - allows an unequivocal evaluation of accuracy. 
As such an analysis is unfeasible for transition metal complexes using today’s classical computers, we selected diatomic systems of both small and large multi-reference character for this purpose. 
Nevertheless, we will demonstrate that it is possible to generalize the results to experimentally more relevant larger molecules.%, including metal complexes. 

The rest of the paper is structured as follows.
In Sect.~\ref{sect:methods}, the computational details are given and  we present the selection of diatomic model systems in Sect.~\ref{sect:model}.
We discuss the effect of orbital choice on the difference between FCI and truncated CC energies in Sect.~\ref{sect:energy}.
As an additional evaluation of the quality of the truncated CC results, we also investigate the orbital dependence of similarity between FCI and CC electron densities in Sect.~\ref{sect:density}. 
Furthermore, to compare the behavior of HF-CC and KS-CC from the practical points of view, we examine the reference dependence of the wave-function diagnostics in Sect.~\ref{sect:diagnostics}, the expected error of the routinely applied approximations in Sect.~\ref{sect:approximation}, and the propagation of molecular energy errors on concrete examples of chemical reactions in Sect.~\ref{sect:reaction}.
The analysis of larger electronic system is found in Sect.~\ref{sect:transition}.
The conclusion, \ref{sect:conclusion}, is followed by the appendix on the FCI invariance of reference.
\begin{table*}[!t]
  \begin{tabular*}{0.99\textwidth}{@{\extracolsep{\fill} } lccccccccc }%| l | c | c | c | c | c|c|c|c|c|}
  \hline 
   \hline 
      Molecule & BH & HF & CF & BF& NO & OF & BN & C$_2$ & B$_2$\\ \hline 
TAE$_{\rm corr}$\% (HF-CCSD) & 97.8 & 95.0 & 89.5 & 89.8 & 89.9 & 88.2 & 83.3 & 84.5 & 75.7 \\ %\hline 
TAE$_{\rm corr}$\%(HF-CCSDT) & 99.8 & 99.7 & 99.5 & 99.3 & 99.0 & 99.1 & 98.0 & 97.8 & 97.0 \\ %\hline 
TAE$_{\rm corr}$\%(HF-CCSD(T)) & 99.8 & 100.1 & 99.8 & 100.1 & 99.6 & 98.5 & 100.3 & 99.6 & 96.9 \\ %\hline 
 \hline 
  \end{tabular*}
    \caption{ Diatomic model systems selected for the present study. As measures of multi-reference character, the correlation contribution to CCSD/CBS, CCSDT/CBS and CCSD(T)/CBS atomization energies are provided in terms of the correlation contribution to CCSDTQP6/CBS total atomization energy, according to the W4-17 database~\cite{karton2017}. Consult the Supplementary Material for the details of the calculations.}
    \label{tab:mr}
\end{table*}

\section{Computational Details}
\label{sect:methods}
All quantum chemical calculations were carried out using the MRCC program (version 2020-02-22)~\cite{kallay2020}, which was linked to the LIBXC density functional library (version 4.3.4.)~\cite{lehtola2018}. Sample input files are found in the Supplementary Material (SM).
Optimized geometry and multiplicity of the studied molecules was taken from previous theoretical works~\cite{karton2017,feldt2019}.
	
\subsection{Electronic energies	}
For all selected model systems, coupled cluster calculations were performed using STO-3G~\cite{stewart1970}, 6-31G~\cite{hehre1972}, cc-pVDZ and cc-pVTZ~\cite{dunning1989} basis sets, at CCS, CCSD, CCSD(T), CCSDT, CCSDTQ and CCSDTQP level of theory. 
Calculations containing even higher order excitations would require unreasonably large computational resources, even for diatomic systems; nevertheless, as pointed out in the next sections, CCSDTQP produces  converged energies of practically FCI quality.
The self-consistent field (SCF) calculations generating the reference orbitals were performed with restricted (RHF/RKS) formalism in the case of singlet molecules and with unrestricted (UHF/UKS) formalism in the case of higher multiplicities. 

Since we aim to draw general conclusions on the effect of Kohn-Sham orbitals on CC accuracy, we selected various types of density functionals for KS-CC calculations. 
Our conception was to pick one functional from each rung of the Jacobs's ladder of KS-DFT methods~\cite{perdew2001}. 
Following this logic of hierarchy, we chose the Hartree~\cite{hartree1928}, LDA~\cite{kohn1965}, BP86~\cite{perdew1986,becke1988}, M06-L~\cite{zhao2006,zhao2008} and B3LYP~\cite{lee1988,stephens1994} formulas as representative examples for the 0th, 1st, 2nd, 3rd and 4th rung, respectively. 
Double hybrid functionals were omitted from the present study as their SCF procedure, hence the resulting set of orbitals, does not differ from that of conventional hybrid functionals. 
However, we added a range-separated hybrid, HSE06~\cite{krukau2006,heyd2003}, as the sixth functional, which although belongs to the 4th rung, shows distinct features from regular hybrids in many aspects. 
Note that CC calculations over range-separated hybrid KS orbitals require a slight modification in the MRCC source code which is described in the SM.

The accuracy of the electronic energy can be easily evaluated by comparing it to the (near-FCI) reference value. 
Still, it must be kept in mind that CC methods are non-variational, which means that a more accurate energy does not necessarily indicate a more accurate electronic wave function. 
Thus, to better establish the role of reference molecular orbitals, we also analyzed their effect on the three-dimensional function of electron density.

\subsection{Electron density}
The unrelaxed electron density~\cite{jensen2006} of a $\Psi_{\rm CC}$ or $\Psi_{\rm KS-CC}$ many-body wave function, expanded in the basis of HF or KS molecular molecular orbitals $\{\varphi(r)\}$, reads
\begin{equation}
    \rho (r) = \sum_{\sigma i j} P^\sigma_{ij} \varphi_{i \sigma} (r) \varphi_{j \sigma} (r)
    \label{eq:dens}
\end{equation}
with $P$ one-electron reduced density matrix (1RDM) where $i$, $j$ and $\sigma$ denote the spatial orbital and the spin index, respectively.

MRCC provides both the 1RDM matrix and the MO information required for the computation of density according to Eq.~\eqref{eq:dens}, i.e., 1RDM  is obtained from the CCDENSITIES output in FCIDUMP format~\cite{knowles1989}, while orbital data are saved as a MOLDEN file~\cite{molden}. 

The density was mapped to a finite grid by Multiwfn~\cite{multiwfn}, whose relevant input, i.e., the data of the corresponding natural orbitals~\cite{jensen2006}, was generated by our hand-coded routine. 
In order to ensure sufficient numerical accuracy, the density was calculated for a grid which was set fine enough to keep the deviation of the integrated electronic charge from the theoretical value below 0.1\%.

\section{Results and Discussion}
\subsection{Selection of model systems}
\label{sect:model}
As the first step of our investigation, we selected a set of diatomic molecules with versatile degrees of multi-reference character. 
The selection was based on the W4-17 dataset where the atomization energy is computed for 200 molecular systems at various HF-CC levels up to sextuple excitations, using complete basis set (CBS) extrapolation~\cite{karton2017}. 
In contrast to $T_1$ or other wave-function diagnostics, the accuracy of total atomization energies (TAE) are investigated as an unambiguous indicator of multi-reference effects. 
In Ref.~\onlinecite{karton2017}, the commonly applied CCSD, CCSD(T) or CCSDT level TAE results were compared to the TAE for CCSDTQP6 of FCI quality.

The reliability of truncated CC methods can be quantified by the proportion of correlation effects taken into account by the calculation. 
The percentage of correlation energy (${\rm TAE_{corr}}\%$) recovered by a particular truncated method $X$ is given by
\begin{equation}
%    TAE_{corr}\% (HF-X) = \frac{TAE(HF-X) - TAE(HF-SCF)}{TAE_{corr}}
    {\rm TAE_{corr}}\% (X) = 100 \frac{ {\rm \Delta  TAE( HF\m } X)}{{\rm \Delta TAE( HF\m CC\m SDTQP6})}\,.
    \label{TAEcorr}
\end{equation}
Here, we introduced the relative TAE for method HF-CC-$X$  which measures the TAE difference obtained for HF-CC-$X$ and for HF-SCF level, i.e., 
\begin{equation}
{\rm \Delta TAE( HF\m}X)= {\rm TAE( HF\m}X)-{\rm TAE( HF\m SCF})\,.
\end{equation}
Eq.~\eqref{TAEcorr} gives 100\% for methods of FCI quality and 0\% for plain Hartree-Fock SCF. 
The typically applied CC methods are expected to give a value between 0\% and 100\%, depending on the CC level and the degree of multi-reference. 
Nevertheless, as the implemented coupled cluster equations are not based on the variational principle, values over 100\%, i.e. the overestimation of correlation effects, are also not excluded. 
We note that the advantageous feature of TAE$_{\rm corr}$\% is that it quantifies the multi-reference character in a system size independent manner, allowing simple comparisons among different molecules.  

As summarized in Tab.~\ref{tab:mr}, nine diatomic systems were chosen for the study (BH, HF, BF, CF, NO, C$_2$, OF, B$_2$, BN), which are all sufficiently small for near-FCI calculations with basis sets of reasonable size. 
According to the TAE$_{\rm corr}$\% covered by lower CC levels, the multi-reference character gradually increases from BH to B$_2$, as ordered from left to right in Tab.~\ref{tab:mr}. The order is primarily based on TAE$_{\rm corr}$\% for CCSDT, but it is mostly consistent with the relation of CCSD and CCSD(T) data. 
The rationale for the reversed order of NO and OF regarding TAE$_{\rm corr}$\%(CCSDT) is the significantly lower TAE$_{\rm corr}$\%(CCSD) and TAE$_{\rm corr}$\%(CCSD(T)) percentages of the latter. 

\subsection{Convergence of HF-CC and KS-CC absolute electronic energies to the FCI limit}
\label{sect:energy}
We began our investigations by computing the electronic energies of the nine molecules in Tab.~\ref{tab:mr} at different HF-CC and KS-CC levels.

In theory, HF-CC and KS-CC converge to the same energy limit, i.e., to the FCI limit for a given molecule and basis set, when gradually increasing the maximal number of excitations, in spite of the fact that considerably different energies might result from HF-CC and KS-CC at the commonly used truncated levels, such as the popular CCSD(T). 
A mathematical proof for the invariance of the FCI limit with respect to the choice of reference orbitals is provided in the Appendix. 
Importantly, once this universal energy limit is determined, the evaluation of the effect of reference orbitals becomes straightforward as the reference enabling the fastest convergence to FCI, i.e., the reference producing the closest energy to FCI at a given CC level, can be considered the most beneficial.

The main issue with the latter approach is that the routinely applied frozen core and density fitting approximations need to be turned off, otherwise the FCI limit is not independent of the reference in practice, which makes any rigorous comparison between HF-CC and KS-CC absolute electronic energies meaningless. 
All-electron coupled cluster calculations without density fitting are, however, very expensive, even for diatomic systems. 
Results of FCI quality can only be obtained with relatively small basis sets: we utilized STO-3G, 6-31G and cc-pVDZ at this stage of the research. 
In the case of all three bases, we managed to perform calculations up to CCSDTQP level, and the deviation between HF-CCSDTQP and KS-CCSDTQP electronic energies was found to be below 0.05 kcal/mol for all molecules. 
Taken this negligibly low value together with the expected negligible contribution of sextuple excitations to the electronic energy (according to the W4-17 dataset, $< 0.1$ kcal/mol change in $TAE$ was observed, even for pathologically multi-reference systems,), it can be stated that CCSDTQP results essentially represent the FCI limit.   
	
The resulting electronic energies are summarized in the Supplementary Material (SM). 
As the size of the basis set was found to have little influence on the relation of HF-CC and KS-CC results, we only present the evaluation of cc-pVDZ calculations in the following whereas the discussion of STO-3G and 6-31G results are found in the SM.

\begin{figure*}[!t]
  \includegraphics[width=0.85\textwidth]{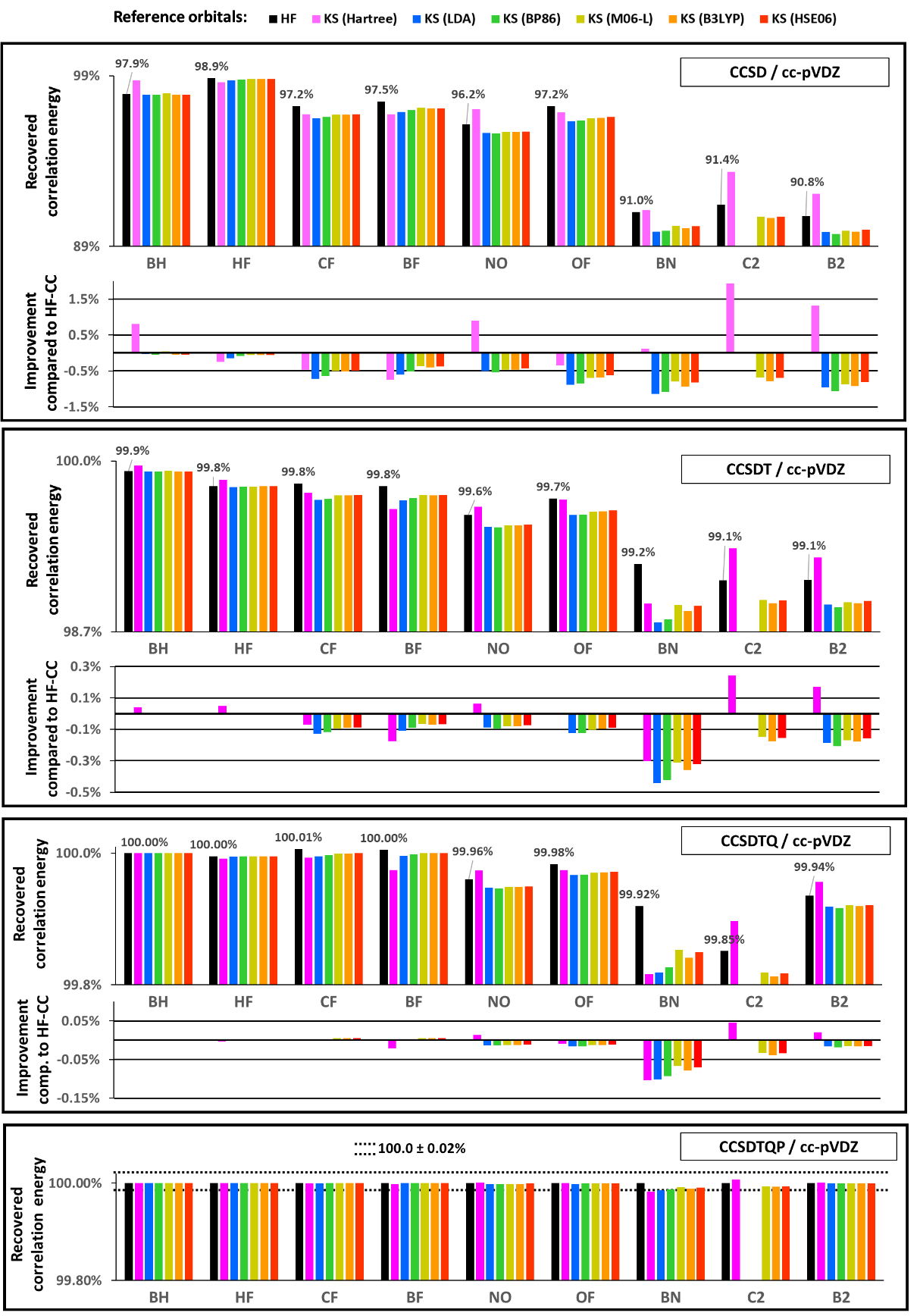}
  \caption{
Effect of the choice of reference orbitals on the accuracy of CC/cc-pVDZ electronic energies, measured against the FCI energy limit (computed at HF-CCSDTQP/cc-pVDZ level in practice).
Results for C$_2$ based on KS-LDA and KS-B686 reference are not presented due to convergence issue.
 \label{fig:energy}}
\end{figure*}

The effect of reference orbitals on the convergence of CC/cc-pVDZ energies to the FCI limit is illustrated in Fig.~\ref{fig:energy}. 
Each of the seven chosen reference orbital sets (generated using the habitual Hartree-Fock method and the six Kohn-Sham density functionals listed in the Computational Details) is assigned to a color, as depicted by the legend on the very top. 
Going from the top black frame to the bottom one of Fig.~\ref{fig:energy}, we present the performance of the increasingly accurate and expensive non-perturbative coupled cluster methods of CCSD, CCSDT, CCSDTQ and CCSDTQP. 
In each frame, the horizontal axis shows the diatomic model systems in the order of increasing multi-reference character (see also Tab.~\ref{tab:mr}). 
On the top vertical axes, the percentage of recovered correlation energy is depicted, which is defined - analogously to Eq.~\ref{TAEcorr}, in a molecule size independent manner - as
\begin{equation}
    E_{corr} \% (X) = 100 \frac{\Delta E(X)}{\Delta E({\rm HF\m CCSDTQP})}
\end{equation}
using $\Delta E(X)=E(X)-E({\rm HF\m SCF})$,
where E stands for the electronic energy and $X$ is the computational method to be characterized. 
Note that this expression gives 100\% for $X$ = HF-CCSDTQP, and should negligibly deviate from 100\% in the case of any KS-CCSDTQP energy, as any of these are practically equal to the FCI limit. 
For sake of brevity, we introduce notation $E_{\rm corr}= \Delta E({\rm HF\m CCSDTQP})$. 
It is notable that all electronic energies, including KS-CC results, were analyzed in terms of the dimensionless $E_{\rm corr} \% (X)$ using the HF referenced electronic energies. 
In this way,  $E(X)$ is the only quantity in the measure  which varies with the reference, accordingly it is ensured  that a higher $E_{\rm corr} \%$  percentage indicates higher accuracy, i. e. smaller deviation from the near-FCI energy.

The convergence to the FCI energy limit upon increasing the CC level can be observed on the change of the scale of the top vertical axes in Fig.~\ref{fig:energy}. At CCSD level (top frame of Fig.~\ref{fig:energy}) 89-99\% of the correlation energy is recovered, depending on the molecule and the reference orbitals. 
The inclusion of higher and higher excitations to the CC ansatz brings the depicted  values closer and closer to the FCI limit of 100\%, i.e., CCSDT: 98.7-100.0\%; CCSDTQ: 99.82-100.01\%. 
At CCSDTQP level, all values reach 100.0\% with a negligible variation of $\pm$0.02\% in $E_{\rm corr} \%$, regardless of the multi-reference character of the molecule and the choice of reference orbitals. 

The bottom vertical axes represent the alteration in the recovered correlation energy percentage upon changing the usual HF reference to KS orbitals, which we defined as
\begin{eqnarray}
    \Delta E_{\rm corr} \% (({\rm KS\m CC}X) &=& |100-E_{\rm corr} \% ({\rm HF\m CC}X)| \\
    &&- |100-E_{\rm corr}\% ({\rm KS\m CC}X)|\,.\nonumber
    \label{eq:deltaE}
\end{eqnarray}
\begin{figure*}[!ht]
  \includegraphics[width=0.85\textwidth]{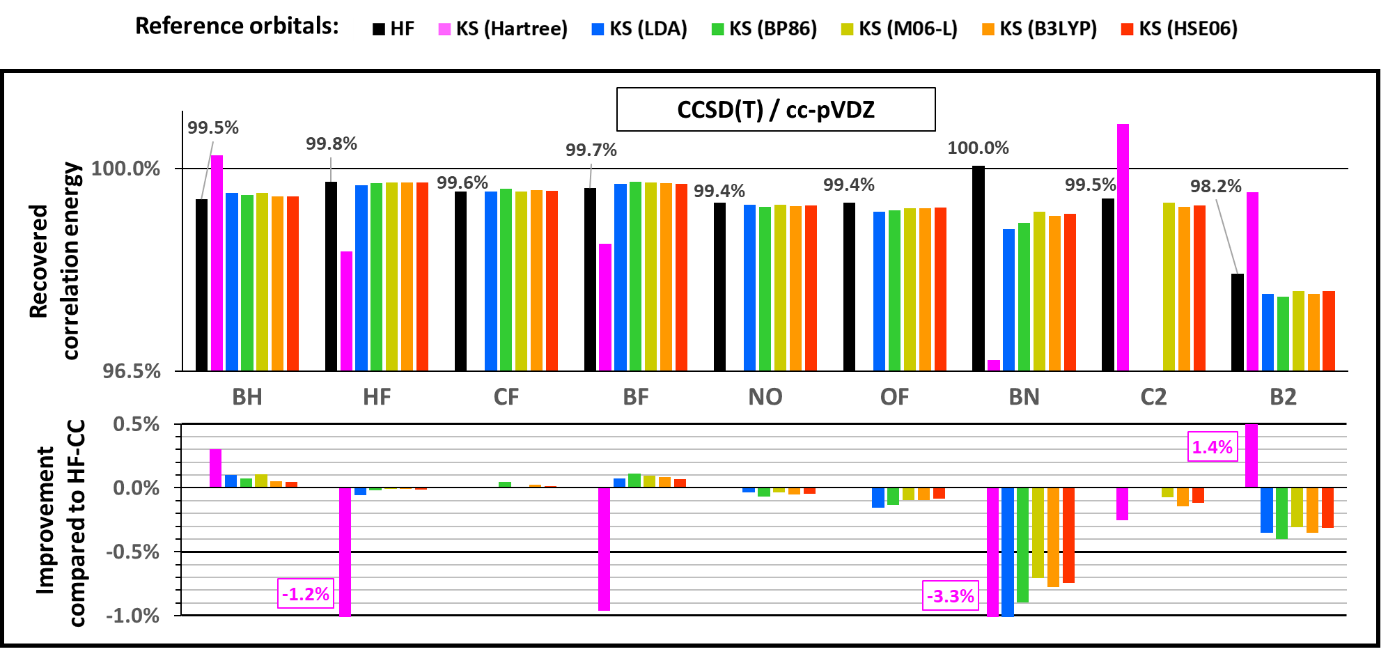}
  \caption{
  Effect of the choice of reference orbitals on the accuracy of CCSD(T)/cc-pVDZ electronic energies, measured against the FCI energy limit (computed at HF-CCSDTQP/cc-pVDZ level in practice). 
  Results for C$_2$ based on KS-LDA and KS-B686 reference as well as for CF, NO, OF based on Hartree reference are not presented due to convergence issue.
 \label{fig:energy_pert}}
\end{figure*}
A positive  $\Delta E_{\rm corr} \%$ indicates the improvement effect of the given KS reference, i.e., $E_{\rm corr}\%({\rm KS\m CC}X)$  approximate more accurately  the FCI limit of 100\%, while a negative value means deterioration relative to the HF-CC$X$ result. 
We note that the absolute values in Eq.~\eqref{eq:deltaE} are introduced due to the non-variational nature of the applied coupled cluster implementation, which might lead not only to the underestimation, but also to the overestimation of correlation energy ($E_{\rm corr} \%>100\%$) at truncated CC levels.

Benchmarks are summarized in Fig.~\ref{fig:energy} which clearly shows   that the change of the habitual HF reference (black bars) to KS (colored bars) yields clear deterioration relative to the FCI limit ($\Delta E_{\rm corr} \%<0$)  for the commonly used functionals. 
In contrast to this,  the most basic Hartree formula (pink bars) provides  an opposite trend for some of the studied molecules and it might facilitate the convergence to FCI ($\Delta E_{\rm corr} \%>0$).
The interesting tendency might be interpreted as the consequence of the intimate relation of the Hartree and the Hartree-Fock functionals which is also witnessed at low orders in the  M{\o}ller-Plesset perturbation theory which is closely connected to the CC expansion~\cite{jensen2006}. 

In particular, at CCSD level (top black frame of Fig.~\ref{fig:energy}), the use of any of LDA, BP86, M06-L, B3LYP and HSE06 references consistently decreases the accuracy of the electronic energy. 
In the following figures molecules are arranged according to their multi-reference character given in Tab.~\ref{tab:mr}.
The deviation from HF-CCSD in the unfavorable direction ranges from a minimal change of 0.0-0.1\% in terms of correlation energy (BH, HF) to up to 0.5\% in moderately multi-reference cases, and up to 1\% in the case of the four highly multi-reference systems. 
Apart from the least correlated BH and HF molecules, the KS-CCSD energies - with the exception of those based on Hartree orbitals - are typically more similar to one another than to the HF-CCSD results, i.e., the difference between the highest and lowest $\Delta E_{\rm corr} \%$ values is generally below 0.2\% and even the highest variations, observed at exceptionally high multi-reference character, BN, B$_2$, do not exceed 0.35\%. 
Note that results for C$_2$ based on KS-LDA and KS-B686 reference are not presented due to convergence issue.
Even though this result is in complete agreement with previous findings~\cite{feldt2019}, it is still a remarkable outcome considering the great variety of density functionals from local density approximation to hybrid formulas. 
We also note that the order of functionals in KS-CC accuracy somewhat reflects the Jacob's ladder of DFT, i.e., LDA from the 1st rung (blue color) and HSE06 from the 4th rung (red color) appear to be the least and most favorable functionals, respectively.
Nevertheless, even the most complex functional produces a less beneficial reference than the HF method. 
In the unique, deviant case of the Hartree functional (pink color), for most of the investigated systems, Hartree reference provides lower KS-CCSD energy than the HF-CCSD, i.e.,  5 molecules out of 9 give positive $\Delta E_{\rm corr} \%$, among which the values of 1.9\% (C$_2$), 1.3\% (B$_2$) and 0.9\% (NO) are the most significant. 
On the other hand, the improvement is not systematic as Hartree orbitals were identified to be even the least accurate in the case of BF  and three other negative $\Delta E_{\rm corr} \%$ values were also found.   
	
Although the inclusion of triple excitations (CCSDT/cc-pVDZ frame of Fig.~\ref{fig:energy}) considerably decreases the variation of electronic energies as expected upon approaching the FCI limit, the above described trends are sustained. 
KS-CCSDT results, which remain practically independent of the functional choice except for Hartree, underperform HF-CCSDT by $0.1-0.3\%$ of the correlation energy, going from low to high multi-reference character. 
In particular, for BH and HF, $\Delta E_{\rm corr}\%$ decreases practically to zero, indicating that CCSDT is of near-FCI quality due to the low extent of correlation.
The beneficial effect of the Hartree functional also appears at CCSDT level for the previously highlighted molecules of C$_2$, B$_2$ and NO ($\Delta E_{\rm corr}\% = +0.06-0.24\%$).
Furthermore, the electronic energy of BH and HF is also improved by Hartree at some extent ($\Delta E_{\rm corr}\% \leq +0.05\%$). 

Taking another step further to CCSDTQ level, the first four molecules of BH, HF, CF and BF are found to reach the convergence of electronic energy, regardless of the reference choice. 
The deviation of KS-CCSDTQ from HF-CCSDTQ is barely detectable ($|\Delta E_{\rm corr}\% |\leq 0.01\%$)  with the sole exception of the Hartree reference in the case of BN.
For NO and the four highest multi-reference cases, the previously noted trends apply, i.e., HF-CCSDTQ provides the closest approximation of the FCI limit, slightly outperforming KS-CCSDTQ by 0.01-0.1\% of $E_{\rm corr}$, with the size of deviation depending on the degree of multi-reference character. 
The effect of orbital choice within the Kohn-Sham framework has negligible influence as only the Hartree functional stands out with some improvements up to $\Delta E_{\rm corr}\% =+0.05\%$ (C$_2$, B$_2$, NO) or, on the contrary, with exceptional deterioration  for BN.

Altogether, it can be concluded that none of the standard KS references is suitable for taking into account the correlation energy part neglected by HF-CC. 
What is more, KS orbitals actually found to slightly decelerate the convergence to the FCI limit.
A systematic deterioration of electronic energy was observed compared to HF-CC, which increases with degree of multi-reference character and amounts up to 1\%, 0.3\% and 0.1\% of $E_{\rm corr}\%$ at CCSD, CCSDT and CCSDTQ level, respectively. 
It should not escape our attention that, given that these proportions also apply to larger molecules such as transition metal complexes, the reference dependent deviation might easily reach multiple kcal/mol units for practically relevant chemical reactions.
  
As the next step, we extended the previously established analyses for the most commonly applied CCSD(T) method presented in Fig.~\ref{fig:energy_pert} where the recovered correlation energy percentages are quite similar to those at CCSDT level.
$\Delta E_{\rm corr}\%$ values, however, range down to -1\%, which better recalls the CCSD results. 
The behavior of high multi-reference systems closely follows the patterns of Fig.~\ref{fig:energy}, i.e., the change from HF to any KS reference results in the consistent deterioration of the electronic energy, the extent of which has little dependence on the density functional. 
The case of low multi-reference character is slightly more complex, as certain systems show distinct CCSD(T) results, i.e., positive $\Delta E_{\rm corr}\%$ values appear with non-Hartree references, ranging up to +0.1\% (for BH and BF). 
Still, the practical significance of this magnitude of improvement is debatable, as it seems to be rather incidental and it is peculiar to two of the least correlated systems with the lowest absolute correlation energy. 
As for the Hartree functional based perturbative results, the calculations are found to be rather unstable, i.e., convergence issues for CF, NO, OF, C$_2$, even the converged energies deviate significantly from the HF-CCSD(T) value.
In summary, according our numerical exploration, the use of typical KS references in CCSD(T) generally provides  energy comparable to the HF counterpart.

\subsection{Convergence of HF-CC and KS-CC electron densities to the FCI limit}
\label{sect:density}
\begin{figure*}[!t]
  \includegraphics[width=0.85\textwidth]{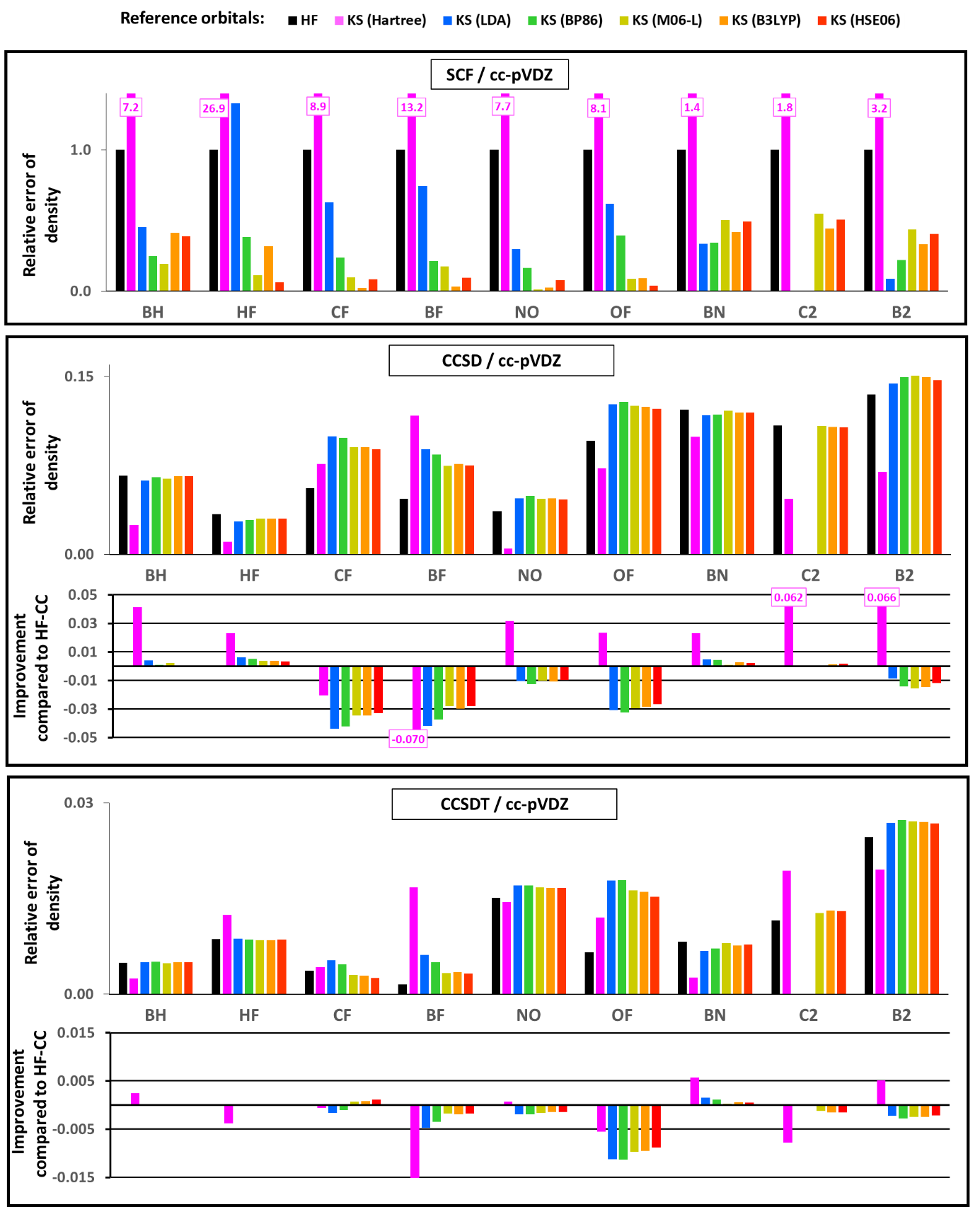}
  \caption{
  Effect of the choice of reference orbitals on the accuracy of CC/cc-pVDZ electron densities. 
  A value of 0 in the relative error corresponds to the full similarity to the FCI density (calculated at HF-CCSDTQP/cc-pVDZ in practice), whereas the value of 1 represents the size of error in HF/cc-pVDZ density, see Eq.~\ref{eq:rho_norm}-\ref{eq:rho_d} in the main text for detailed mathematical definitions. 
  Results for C$_2$ based on KS-LDA and KS-B686 reference are not presented due to convergence issue. 
 \label{fig:density}}
\end{figure*}
\begin{figure*}[!t]
  \includegraphics[width=0.85\textwidth]{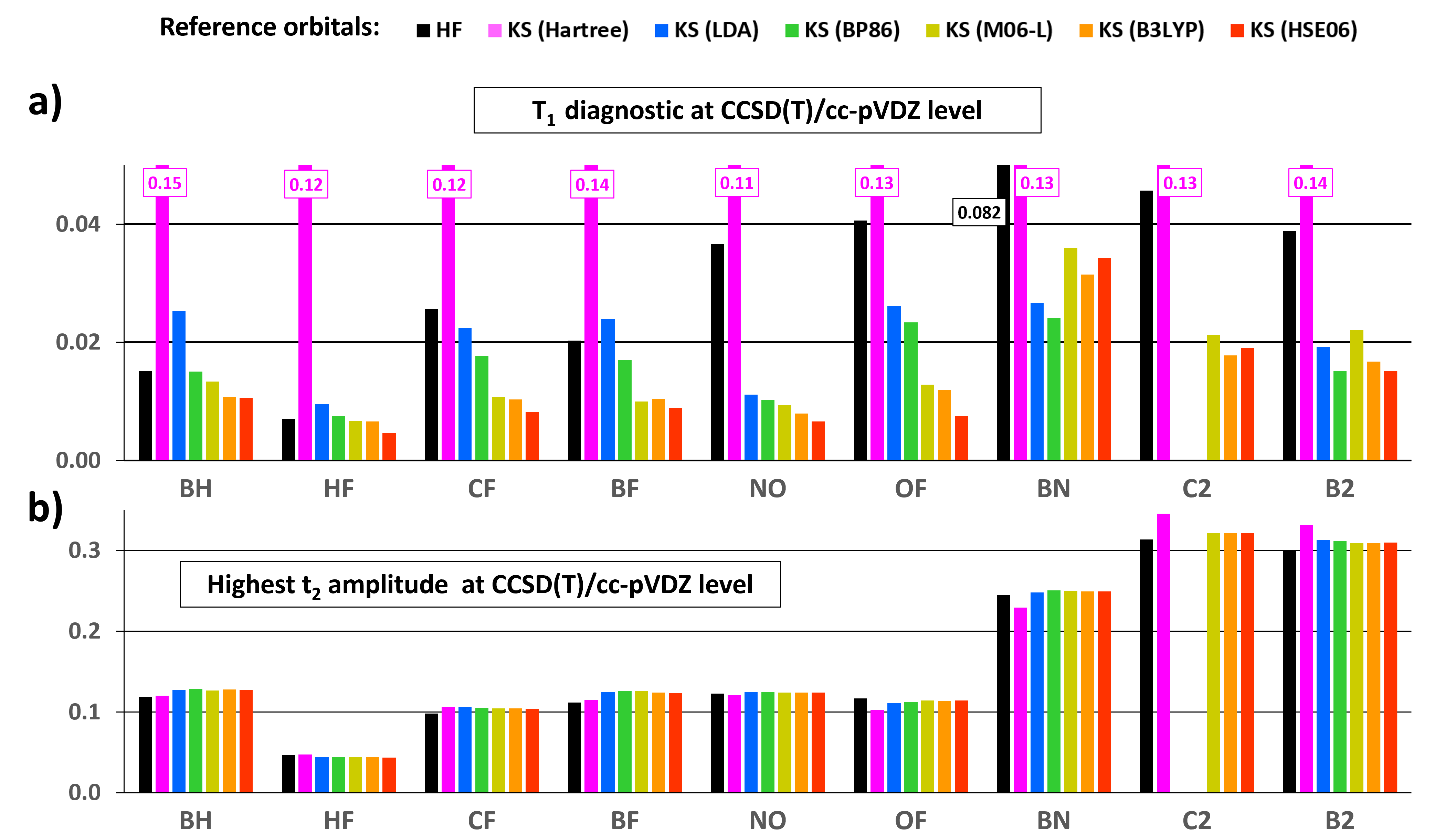}
  \caption{
  Effect of the choice of reference orbitals on the coupled cluster wave-function diagnostics at CCSD(T)/cc-pVDZ level. (a) $T_1$ diagnostic; (b) largest double  amplitude, max$(|t_2|)$. Results for C$_2$ based on KS-LDA and KS-B686 reference are not presented due to convergence issue.    
 \label{fig:diagnostic}}
\end{figure*}

It is notable that the applied CC approach is not a variational approach hence lower CC energy does not necessarily indicate a more accurate description of the problem. Therefore, in the following, we attempt to assess the accuracy of the computations in an alternative way, i.e.,  we benchmark the precision of the CC calculations in terms of the associated electron density instead of the numerically demanding direct comparison of  the corresponding many-body wave functions. 
The application of this analysis is also motivated by getting physical insight on the effect of using various correlation-exchange functionals as we not only monitor densities obtained for the coupled cluster excitation levels but for Kohn-Sham theory as well. 

In practice, densities to be compared are evaluated on the same finite grid. In general, the similarity of two vectors, such as electron distributions  $\rho$ and $\rho'$, can be quantified by various measures, e.g., cosine similarity, relative entropies, Minkowski distances. 
In this paper, we apply the standard Euclidean distance,  i.e., the norm of the difference of the vectors, defined as
\begin{equation}
    \| \rho - \rho' \| = \sqrt{\sum ( \rho_i - \rho_i' )^2}
    \label{eq:rho_norm}
\end{equation}
where index $i$ runs over the grid. 
Similarly to the energy based analysis presented in the \ref{sect:energy}, HF-CCSDTQP level density result serves as reference of practically FCI quality in the following CC density benchmarks. 
Accordingly,
\begin{equation}
    D(X) = \| \rho(X) - \rho({\rm HF\m CCSDTQP}) \|
    \label{eq:rho_D}
\end{equation}
measures the deviation of density obtained for $X$ computational model from the reference result. The relative error of method $X$ is defined as 
\begin{equation}
    d(X) = \frac{D(X)}{D({\rm HF\m SCF})}
    \label{eq:rho_d}
\end{equation}
which measures the error in units of $D$(HF-SCF). 
By definition, $d(X)$ is exactly 1 and 0 for $X$=HF-SCF and $X$=HF-CCSDTQP,  respectively.  
For general $X$ cases, where the corresponding wave function recovers a significant portion of the correlation effects, $d(X)$ tends towards 0.   
The numerical data of $d(X)$ is presented in Fig.~\ref{fig:density} for various excitation levels. 

Comparing SCF level error measures, we find that the high-level Kohn-Sham theories provide typically the most accurate single-determinant approximation, i.e., the corresponding errors are  a small fraction of the HF-SCF error. 
Interestingly, molecules of intermediate multi-reference character are found to be represented the most faithfully by the high-level KS-SCF theories. 
Also in agreement with the expectations, the error of the solution  Hartree-SCF, which neglects any exchange and correlation effects, is outstandingly large, i.e., it is typically multiple of  $D$(HF-SCF).

For increasing CC excitation levels, the tendencies of electron-density errors  are generally in line with the conclusions  of the correlation-energy analysis detailed in Sect.~\ref{sect:energy}. 
Most importantly, the lowest deviations are typically found for CC results of Hartree and Hartree-Fock references, i.e. the addition of computationally rather cheap CCSD excitations already compensate their limitations found at SCF level of theory. 
Similarly to the observations found for CC energies, the deviation of electron-density error is quite even  for the various KS functionals compared to tendencies found for the aforementioned correlation-free references. 
In line with expectations corroborated by the findings for molecular energy, low-level CC theory also shows an overall tendency to provide more accurate electron density for single-reference molecules.
Also remarkable that low-level CC theory applied on high-level functionals (B3LYP and HS06)  can be even counterproductive for the moderately correlated problems, i.e, their extraordinarily accurate SCF electron density is slightly deteriorated at KS-CCSD level of theory.

For CCSDTQP level of theory, the error of the wave functions is already found to be practically zero confirming the convergence of the high-level CC calculations to the exact solution.

\subsection{The relation of $T_1$ and max($|t_2|$) diagnostics to the accuracy of electronic energy}
\label{sect:diagnostics}
The $T_1$ diagnostic of Eq.~\eqref{t1} and the $\max(|t_2|$) value, i.e., the largest absolute value among doubles amplitudes, are commonly applied in the literature to check the reliability of coupled cluster calculations. 
Thus, we were naturally interested in how the reference choice affects these wave-function diagnostics and whether the accuracy trends described in the previous sections (especially the outstanding performance of HF-CC) are reflected in $T_1$ and $\max(|t_2|)$.

Fig.~\ref{fig:diagnostic}a shows the  $T_1$ diagnostic for the diatomic model systems at CCSD(T)/cc-pVDZ level of theory. 
The magnitude of singles CC amplitudes has remarkable reference dependence, i.e, in most cases, $T_1$ value for HF based result are drastically larger relative to KS counterparts. 
We find that all types of non-Hartree density functionals typically decrease the $T_1$ diagnostic except for a minor number of increases with LDA and BP86 functionals in the case of the least multi-reference BH, HF and BF systems. 
Furthermore, with the exception of BN and B$_2$, the decreasing order of $T_1$ values from LDA to HSE06 fairly corresponds to the Jacob's ladder hierarchy. 
The most conspicuous example is probably OF, where the HF-based $T_1$ diagnostic of 0.04 is already reduced to 0.026 when using a simple LDA functional, and is decreased to as low as 0.007 with HSE06.

These findings are in line with the previous KS-CC studies reporting decreased $T_1$ values relative to HF-CC~\cite{harvey2003,vasiliu2015,fang2016}. 
Furthermore, we also observe an overall trend of increased $T_1$ values  for the more correlated problems in the case of the common references.
In contrast to this trend, in the case of CC solutions based on the Hartree reference, which completely neglects quantum exchange and correlation effects, the diagnostic produces exceedingly large value, i.e., 0.11-0.15, regardless of the particular molecule.

The tendencies found for $T_1$ might be interpreted as the footprint of orbital relaxation effect induced by single excitations, i.e.,  larger $T_1$ is detected for the less accurate single-determinant approximations of Hartree-Fock and -in particular- Hartree theories.
This argument is corroborated by our systematic numerical investigation of energy and density, see Sect.~\ref{sect:energy} and \ref{sect:density}, which also shows that lower $T_1$ values do not indicate more accurate approximation of the molecule. 

The reference dependence of the other widely used wave-function diagnostic, $\max(|t_2|)$, is plotted in Fig.~\ref{fig:diagnostic}b. 
In contrast to $T_1$ diagnostic, $\max(|t_2|)$ is found to be practically independent of the reference, with a variation below 0.01 in most cases. 
Even the Hartree functional, which otherwise always shows distinctive behavior, fits the rest of the data points, though it slightly stands out at high multi-reference character.
It is also notable that $\max(|t_2|)$ systematically increases for problems of strongly correlated character, see the outstandingly $\max(|t_2|)$ values of BN, C$_2$ and B$_2$. 

To conclude on the use of CC-amplitude based diagnostics, according to our numerical exploration, $\max(|t_2|)$ - rather than $T_1$ - could be applied to estimate the degree of multi-reference character, hence the reliability of lower-level truncated CC calculations. 
Nevertheless, we found no obvious relation between the reference dependent deviations in the diagnostic value and the relative accuracy of HF-CC and different KS-CC methods reflected by energy and electron density. 
\begin{figure*}[!t]
  \includegraphics[width=0.85\textwidth]{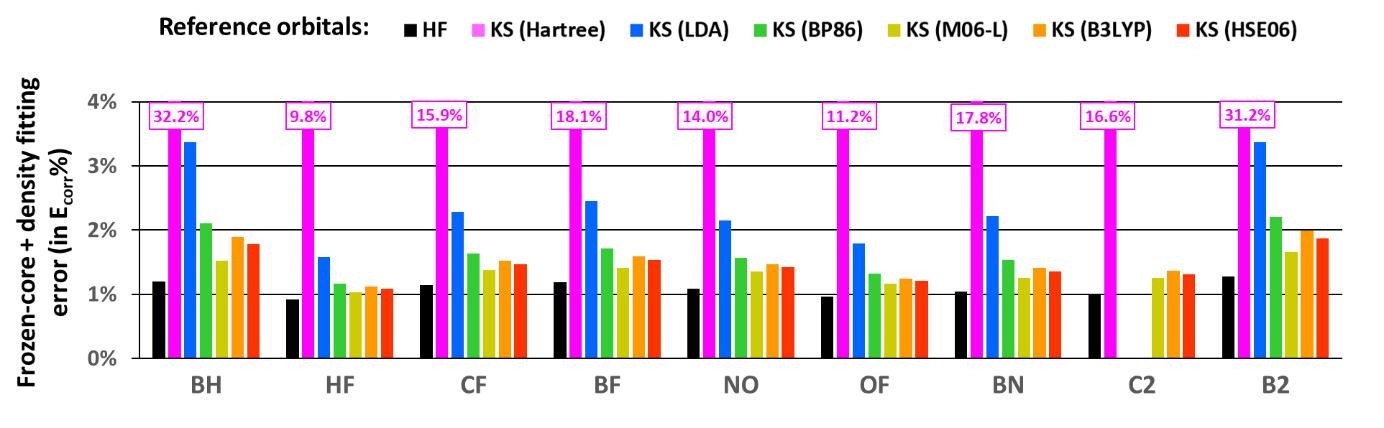}
  \caption{
  Effect of the choice of reference orbitals on the size of error in absolute electronic energy deriving from frozen core and density fitting approximations, computed at CCSDTQP/cc-pVDZ level. The error is given in terms of percentage of correlation energy. Results for C$_2$ based on KS-LDA and KS-B686 reference are not presented due to convergence issue. 
 \label{fig:frozencore}}
\end{figure*}

\subsection{Effects of frozen core and density fitting approximations  }
\label{sect:approximation}
Until this point, we discussed coupled cluster calculations without any additional approximations, in order to focus purely on the effects of reference choice and exclude any other factors. 
In practice, however,  frozen core approximation (FCA) and density fitting (DF) are commonly applied to speed up calculations, which introduce  additional sources of error in the electronic energy. 
In FCA the core orbitals are kept doubly occupied in the CC calculations, thus the number of active orbitals can be significantly reduced for heavy elements by the negligible loss of correlation effects of the low-lying core electrons. 
DF is used to provide computationally cheap approximation for numerical integrals, correspondingly, we found that the  error introduced by DF is detectable but rather marginal compared to the chemical accuracy scale.
  
To investigate the magnitude and the reference dependence of these typical error sources, we repeated the energy calculations described in \ref{sect:energy} applying both FCA and DF.
The obtained CC electronic energies are found in the Supplementary Material.

As already discussed in \ref{sect:energy}, the use of FCA and DF breaks down the theoretically established reference invariance of the FCI energy; thus, we firstly examined how the reference affects the approximated CCSDTQP (practically FCI) level energy. 
In Fig.~\ref{fig:frozencore}, we plotted the deviation of these energies from those computed at non-approximated CCSDTQP level, expressing it in terms of the percentage of the exact correlation energy, $E_{\rm corr}$. 
	
Fig.~\ref{fig:frozencore} clearly shows that the Hartree-Fock reference comes with the smallest error, which amounts approximately 1\% of the correlation energy for all molecules. 
As for Kohn-Sham referenced methods, the error of approximations consistently follows the order of M06-L $<$ HSE06 $<$ B3LYP $<$ BP86 $<$ LDA $<$ Hartree. 
The first four functionals in the latter order, which produce errors between 1\% and 2\% of $E_{\rm corr}$, show very similar behavior, in the sense that their approximated FCI limits are closer to each other than to that of HF-CC. 
In the case of LDA, the use of frozen core and density fitting causes   2\% and 4\% error.
 Most notably, due to the complete neglect of exchange processes of core electrons,  10-30\% portion of $E_{\rm corr}$ is lost with Hartree references which also implies that the application of Hartree orbitals is  severely limited for small electronic systems. 
 
Of course, the magnitude of the error is a function of the CC level, but we find that the truncation of the CC ansatz leaves the results shown in Fig.~\ref{fig:frozencore} practically unchanged, see SM for the error of approximations with CCSD, CCSD(T) and CCSDT.
That is, the same amount of electronic energy (as given in Fig.~\ref{fig:frozencore}) is neglected by frozen core and density fitting at all CC levels, implying that the slower convergence of KS-CC methods to the FCI limit, as established in \ref{sect:energy}, also holds true for approximated CC. 
This tendency also reveals that the contribution of the cores is rather marginal in the overall correlation effects  validating the applicability of the FCA.
	
Altogether, our results indicate that the HF-CC methods not only converge faster to its FCI, but also have more accurate FCI limit if the common approximations are taken into account. 
The consistent performance of the HF reference is rather natural as these orbitals are optimized for describing exchange effects which can be also captured in the FCA at SCF level of theory.

\subsection{Implications on the accuracy of reaction energies: error propagation}
\label{sect:reaction}
\begin{figure*}[!t]
\includegraphics[width=0.85\textwidth]{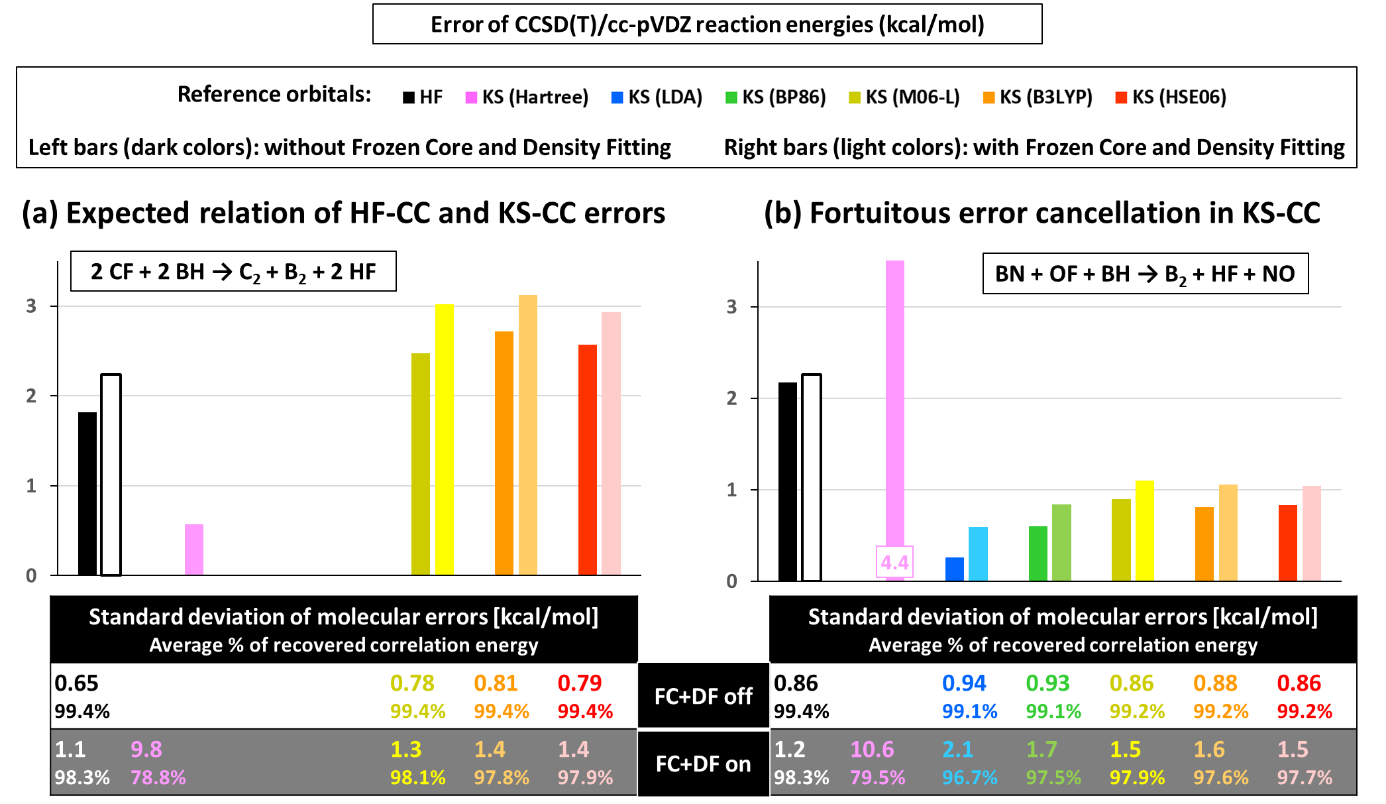}
\caption{
Effect of the choice of reference orbitals on the error in CCSD(T)/cc-pVDZ reaction energies (kcal/mol), determined against non-approximated CCSDTQP/cc-pVDZ (practically FCI/cc-pVDZ) data. 
Two hypothetical chemical reactions among the studied diatomic model systems ((a) 2CF + 2BH $\rightarrow$ C$_2$ + B$_2$ + 2HF; (b) BN + OF + BH $\rightarrow$ B$_2$ + HF + NO) are presented as representative examples. 
The left-side (dark colored) bar and the right-side (light colored) bar belonging to each reference orbital set show the error of the non-approximated calculation and that of the approximated calculation (with frozen core (FC) and density fitting (DF)), respectively. 
Tables below discuss the origin of reaction energy errors at the level of molecular CC electronic energies: provided are the standard deviation of individual molecular errors, i.e., differences between molecular CCSD(T) and FCI electronic energies in kcal/mol, and the average of the percentages of recovered molecular correlation energy. 
Missing data is not presented due to convergence issue of molecular calculations. 
Graph (a)  shows the generally expected behavior, i.e.,  larger deviation of molecular errors for typical KS-CC yields larger reaction energy error.
Graph (b) illustrates the case of an accidental error cancellation, i.e., smaller reaction energy errors in spite of the slightly larger variety of molecular errors for KS-CC.
\label{fig:reaction}}
\end{figure*}
As we have seen in the previous sections, the use of Hartree-Fock reference orbitals in truncated CC calculations is expected to lead to the closest approximation of the FCI limit of the molecular electronic energy $E$. 
In quantum chemical studies, however, we are typically interested in the accuracy of reaction energies, which are calculated as the difference of the algebraic sum of the individual $E$  energy of the product molecules and the reactants.

Although it could be argued at first sight that the substitution of the most reliable HF-CC energies necessarily provides the most accurate results, this seemingly reasonable statement is not strictly true. 
Note that the variation – rather than the magnitude – of molecular errors determines the accuracy of the reaction energy.
As KS-CC energies are spread in a  wider range compared to HF-CC results, the chance for less favorable reaction energy error for HF-CC is expected to be relatively low.
Nevertheless, it is possible that a fortuitous cancellation of molecular errors results in the outperformance of KS-CC compared to HF-CC.

To demonstrate the above reasoning on the example of the popular CCSD(T) method, we presented the accuracy of CCSD(T)/cc-pVDZ reaction energies of two hypothetical reactions among the diatomic model systems. 
Fig.~\ref{fig:reaction}a shows the reference dependence of CCSD(T) error in kcal/mol for the reaction of 2CF + 2BH $\rightarrow$ C$_2$ + B$_2$ + 2HF, which follows the expected pattern, i.e., KS-CC produces approximately 1 kcal/mol higher errors than HF-CC, regardless of the functional applied. 
Frozen core and density fitting approximations (see light-colored bars for each reference) slightly enhance the error with all references but have negligible influence on the relation of accuracies. 
As shown by the table below the bar graph, the enhanced uncertainty of KS-CC can be traced back to the larger variation of molecular CCSD(T) errors.
Namely, upon replacing the HF orbitals to KS reference, the standard deviation of errors increases from 0.65 to 0.8 and from 1.1 to 1.3-1.4 in the case of all-electron and frozen-core calculations, respectively. 
We also note that the standard deviation correlates with the overall accuracy of molecular electronic energies (represented by the average of percentages of recovered molecular correlation energy $E_{\rm corr}\%$ in the table): as suggested above, the closer the energies are located the FCI limit (100\%), the smaller the deviation is. 
The only ambiguous result in Fig.~\ref{fig:reaction}a is the small error of the Hartree functional with FC and DF approximations ($\sim$0.5 kcal/mol).
As this method is unequivocally inaccurate, i.e., only 78.8\% of $E_{\rm corr}$ is taken into account by CCSD(T), and its performance also varies drastically among molecules (standard deviation: 9.8 kcal/mol), it is clear that a case of fortunate error cancellation was found.

While the majority of chemical reactions involving the nine diatomic species shows similar behavior to Fig.~\ref{fig:reaction}a, consult SM for further examples, some exceptions are also identified  which understood by fortuitous error cancellation. 
Fig.~\ref{fig:reaction}b shows the reaction energy of BN + OF + BH $\rightarrow$ B$_2$ + HF + NO, where most KS-CCs reaction energy errors halves  the  HF-CC counterpart.
The low KS-CC reaction energy errors are unexpected regarding both the decreased average $E_{\rm corr}\%$ values and the slightly enhanced standard deviations of molecular errors compared to HF-CC. 
We also note the outstandingly high reaction energy error (4.4 kcal/mol) of the Hartree functional, which is in this case fully consistent with the enormous variation of individual errors.
The high consistency of KS-CC results is to be investigated in a subsequent work.

To summarize the above analysis, higher precision in calculating the molecular electronic energies does not guarantee superior performance in predicting reaction energies. 
Nevertheless, the tested reactions confirm the expectations in general, i.e., we find that the reaction energy is tendentiously more accurate for HF-CC relative to the analogous KS-CC results as the consequence of its lower variation of errors in molecular energies.  
It is also notable that incidental outperformance of KS-CC owing to error cancellations can be observed in some cases. 

\begin{figure}[!t]
\includegraphics[width=0.48\textwidth]{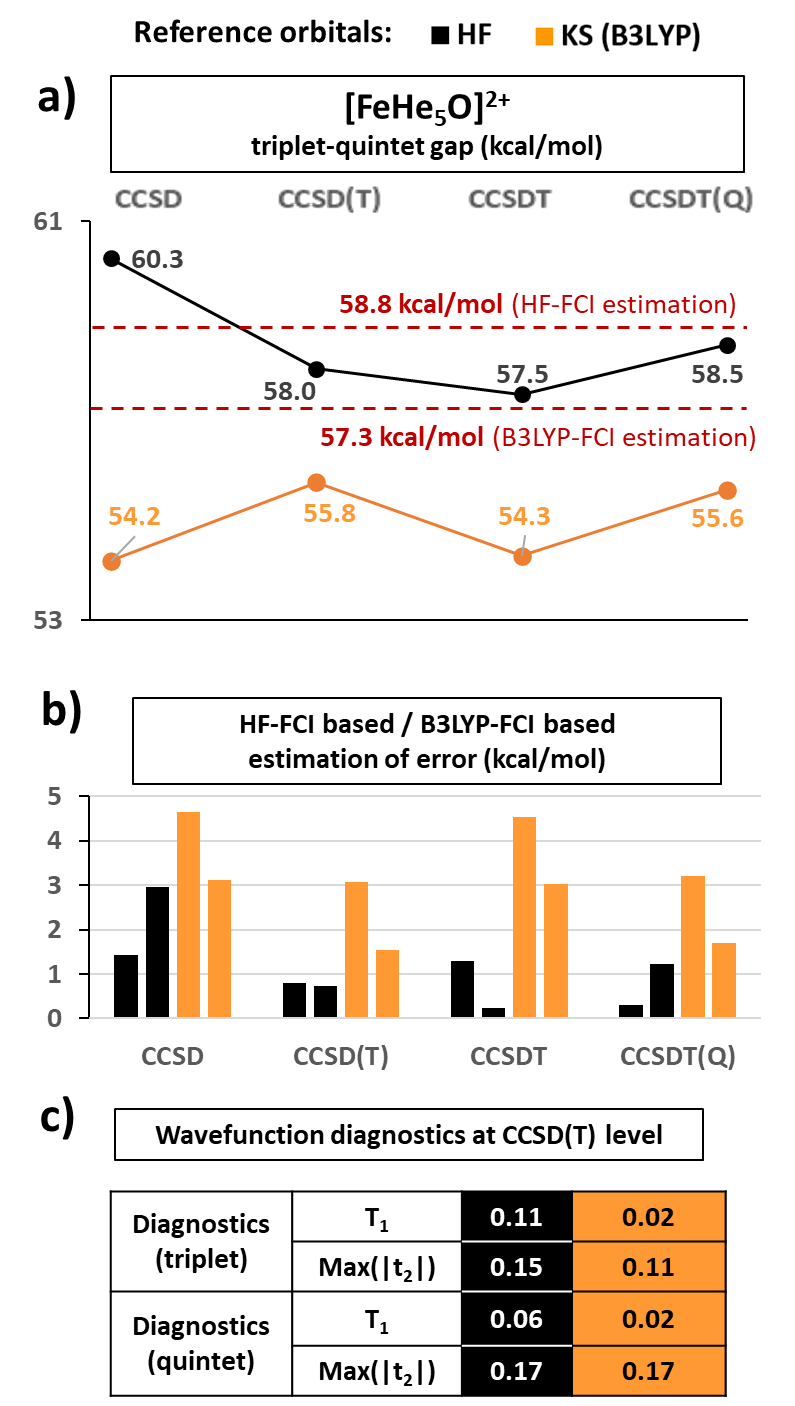}
\caption{
Analysis of the triplet and quintet states of the helium model of [FeHe$_5$O]$^{2+}$:
a) Energy gap computed at truncated CC models with FCA+DF.
b) Error analysis of the gap.
c) Wave-function diagnostics using $T_1$ and max($|t_2|$) measures.
\label{fig:transition}}
\end{figure}

\subsection{Can the results be generalized to larger systems? – Example of a practically relevant transition metal complex}
\label{sect:transition}
Given our results obtained for diatomic model systems, it is naturally of interest whether similar trends are also valid for larger molecules including transition metal (TM) complexes. 
Thus, we searched for a practically relevant complex, which is however small enough to perform calculations of closely FCI quality. 
After deliberation and performing extensive tests on the limits of computational performance, we selected the helium model of Harvey~\cite{feldt2019}. 
This tetragonal bipyramidal complex, structured as [FeHe$_5$O]$^{2+}$, is constructed in a way to resemble the spin density of C-H bond activating iron-oxo catalysts. 

Herein, we examine the effect of reference orbital choice on the triplet-quintet gap of [FeHe$_5$O]$^{2+}$. 
Considering basis sets of reasonable size, such as cc-pVDZ, all-electron CCSDTQP calculations, used as near-FCI reference for diatomic molecules, would be unfeasible even for such a small TM complex. 
Nevertheless, it is possible to calculate the electronic energy of triplet and quintet states at FCA-CCSDT(Q)/cc-pVDZ level with frozen core and density fitting approximations. 

Furthermore,  corresponding to findings of Sect.~\ref{sect:approximation}, we attempted to approximate all-electron CCSDT(Q)/cc-pVDZ level theory
as
\begin{equation}
    E{\rm(CCSDT(Q)} \approx E{\rm(FCA\m CCSDT(Q)} + \Delta E_{\rm FCA}\,,
    \label{eq:approx}
\end{equation}
i.e., the error of FCA and DF can be already estimated at low CC levels, such as $\Delta E_{\rm FCA}=E$(FCA-CCCSD(T))$-E$(CCSD(T)).
Finding that the approximated energy of Eq.~\ref{eq:approx}   is nearly independent of the orbital choice for the investigated TM model, the numerical result can be considered of near-FCI quality.

More specifically, the difference between HF and B3LYP referenced triplet-quintet gaps obtained in this manner ($E_{\rm gap}{\rm (HF)}= 58.8$  kcal/mol and $E_{\rm gap}{\rm (KS\m B3LYP)}= 57.3$ kcal/mol, respectively) is below 1.5 kcal/mol, which approaches the requirement of chemical accuracy and also indicates close convergence to the FCI limit. 
We note that only the popular B3LYP functional was tested here among the KS-CC methods due to the significant computational cost. Additionally, the computed gaps  deviate from the values presented in ref.~\onlinecite{feldt2019}  due to the application of different basis sets.

Fig.~\ref{fig:transition}a visualizes the effect of orbital choice on the triplet-quintet gap at  truncated levels of CCSD, CCSD(T), CCSDT and CCSDT(Q) with FCA+DF approximation. 
Though the exact value of the FCI limit is uncertain (due to the 1.5 kcal/mol difference of the two aforementioned estimations shown as red dashed lines), the performance of HF-CC seems superior to B3LYP-CC according to the following analysis. 

In Fig.~\ref{fig:transition}b the error of the computational models, i.e., black bars for HF-CC and orange bars for B3LYP-CC, is estimated relative to  both $E_{\rm gap}{\rm (HF)}$ and $E_{\rm gap}{\rm (KS\m B3LYP)}$ reference values in  left and right bars, respectively.
It is notable that lower errors are obtained for HF-CC approach regardless of FCI estimation. 
The deterioration of the results by the B3LYP reference is attributed partly to the slower convergence of B3LYP-CC electronic energies to the FCI limit and partly to the larger frozen-core error of B3LYP-CC (see SM for details), which is in complete agreement with our observations on smaller diatomic systems. 

Moreover, as summarized in Fig.~\ref{fig:transition}c, the reference dependence of wave-function diagnostics follows the trends described in Sect.~\ref{sect:diagnostics}, i.e., the $T_1$ diagnostic for B3LYP-CCSD(T) is considerably decreased compared to HF-CCSD(T), while max($|t_2|$) remains essentially constant. 

Altogether, these findings suggest that the conclusions drawn for the nine diatomic molecules are generally applicable for KS-CC.

\section{Conclusion}
\label{sect:conclusion}
The present computational study investigated the effect of orbital choice on the accuracy of coupled cluster calculations compared to reference calculations of FCI quality for the first time to the best of our knowledge. 
According to our numerical exploration, the following points should be taken into consideration in future CC studies involving Kohn-Sham reference orbitals.

\begin{enumerate}
\item KS-CC calculations, regardless of the  exchange-correlation density functional used for the generation of reference orbitals, produce less accurate molecular electronic energies and densities (relative to the FCI result) than the habitual HF-CC approach with equal CC level and basis set. 
Although HF-CC and KS-CC all-electron methods converge to the same FCI limit upon increasing the CC level, this convergence occurs somewhat slower with KS common references. 
What is more, the error of the routinely applied frozen core and density fitting approximations is larger for KS-CC, which further enhances the difference from HF-CC in accuracy, in favor of the latter methodology.
    
\item  The $T_1$ diagnostic of KS-CC wave functions of typical functionals is considerably lower than that of the corresponding HF-CC wave functions in line with literature findings. 
Nevertheless, this seemingly favorable decrease in $T_1$ is neither corroborated by the improvement of correlation energy, nor by the accuracy of the corresponding density. 
On the other hand, the largest $t_2$ amplitudes seem as a reference-free indicator of multi-reference character.
    
\item In spite of the higher errors in individual molecular electronic energies,  KS-CC might still outperform HF-CC in the accuracy of reaction energies corroborating previous observations as a result of cancellation of errors. 
However, the better performance of HF-CC in absolute energy indicate that HF-CC is expected to provide more reliable energy differences in general.
\end{enumerate} 
   		
Although the application of KS-CC might seem unfavorable considering the above points, we emphasize that this methodology is a very useful alternative of HF-CC in case convergence issues encountered in HF-SCF or HF-CC calculations and also because our calculated KS-CC reference energies does not indicate significant differences in absolute energy. 
Additionally, at sufficiently high CC level, determined by the multi-reference character of the studied molecule, the difference between HF-CC and KS-CC results disappears in the calculations and both can be reliably applied.
Furthermore, the relatively small variation of errors using truncated KS-CC with different references of the Jacob's ladder suggest that already computationally cheap functionals, e.g., BP86, might be considered competitive.
 
Finally, our results also imply that alternative post-HF methods approaching the FCI limit become less sensitive to the underlying orbital set. Frozen core approximation introduces some additional error in KS-based molecular energies but it is expected to cause a diminishing effect for light elements in estimating vertical electronic excitations.

\section*{Acknowledgements}
Discussions with \'Ad\'am Gali and P\'eter R. Nagy as well as support from the National Research, Development and Innovation Office (NKFIH FK-20-135496) are greatly acknowledged. 
P.T. and G.B. thank the support from the Wigner Student Scholarship of the Wigner Research Centre for Physics and the Bolyai Research Scholarship of the Hungarian Academy of Sciences, respectively. 
We acknowledge KIF\"U for awarding us access to computational resources based in Hungary.  
Z.B., T.S., and G.B. would like to thank the University of Alabama and the Office of Information Technology for providing high performance computing resources and support that have contributed to these research results. 
This work was made possible in part by a grant of high-performance computing resources and technical support from the Alabama Supercomputer Authority.

Dedicated to the loving memory of professors G\'eza Tichy, J\'anos Pipek and Gyula Radnay.

%\newpage
\section*{Appendix}
In the following we give an illustrative mathematical derivation for the invariance of the FCI solution with respect to the choice of reference orbitals.

Let us consider two different sets of reference orbitals, e.g., a set of Hartree-Fock and a set of Kohn-Sham molecular orbitals  which will be referred to as $\{\varphi'\}$ and $\{\varphi''\}$ in the following, respectively. 
Considering that analogous formula are derived for both orbital sets, for sake of compactness, double notation is introduced  in the following derivations, e.g., $\varphi'^{(\prime)}$ denotes both $\varphi'$ and $\varphi''$ simultaneously.

The basis consisting of $m$  basis functions $\{\chi\}$  is used to expand the molecular orbitals as
\begin{equation}
    \varphi_i'^{(\prime)} = \sum_{j=1}^m c_{ij}'^{(\prime)} \chi_j\,.
    \label{eq:MO}
\end{equation}

Note that the HF and KS procedure creates $m$  molecular orbitals (MOs) from the $m$ atomic basis functions. 
However,  constructing a Slater determinant $\Phi'^{(\prime)}$, only $n$ MOs, corresponding to the number of electrons, are selected out of $m$ MOs. 
The $k$th Slater determinant of the wave function ansatz, containing the $k_1$th, $k_2$th, $\dots$, $k_n$th orbitals reads as 
\begin{equation}
    \Phi_k'^{(\prime)} = \hat{A} \left(\prod_{i=1}^n\varphi_{k_i}'^{(\prime)} (i)\right)\,,~
    \label{eq:SD}
  \end{equation}
where $\hat{A}$ is the antisymmetrizing operator and the parentheses indicate that each one-electron orbital contains the spatial coordinates of a single electron as variable. 
There are $\binom{m}{n}$ possible ways of orbital selection, i.e, the order among orbitals does not make a difference because $\hat{A}$ permutes them anyway, thus FCI ansatz $\Psi'^{(\prime)}$ is expanded in the basis of $d=\binom{m}{n}$ physically relevant Slater determinants, 
\begin{equation}
    \Psi'^{(\prime)} = \sum_{k=1}^{d} C_k'^{(\prime)} \Phi_k'^{(\prime)}\,.
    \label{eq:FCI}
\end{equation}

The Slater determinants in Eq.~\ref{eq:FCI} are expressed in terms of the basis functions using Eq.~\ref{eq:MO} and Eq.~\ref{eq:SD} as

\begin{equation}
    \Psi'^{(\prime)} = \sum_{k=1}^{d} C_k'^{(\prime)} \hat{A} 
    \left(\prod_{i=1}^n \left( \sum_{j=1}^m c_{k_ij}'^{(\prime)} \chi_j (i) \right) 
 \right)\,.
    \label{eq:FCI-MO1}
\end{equation}

By expanding the products of Eq.~\ref{eq:FCI-MO1},  the Slater determinants can be rewritten as a linear combination of $n$-factored products of basis functions, i.e.,	
\begin{equation}
    \Psi'^{(\prime)} = \sum_{k=1}^{d} C_k'^{(\prime)} \sum_S w_{Sk}'^{(\prime)} \prod_{i=1}^n\chi_{S_i} (i)\,,
\end{equation}
where $S$ denotes a selection scenario.
By interchanging the order of summations, the coefficients of basis  product $\prod_{i=1}^n\chi_{S_i} (i)$ corresponding to selection $S$ can be merged into a single variable, i.e.,
\begin{equation}
    W_{S}'^{(\prime)} =  \sum_{k=1}^{d} C_k'^{(\prime)} w_{Sk}'^{(\prime)}\,.
    \label{eq:W}
\end{equation}
Finally a simple expression for the many-body wave function is written as  
\begin{equation}
    \Psi'^{(\prime)} = \sum_S W_{S}'^{(\prime)}\prod_{i=1}^n\chi_{S_i} (i) \,.
    \label{eq:FCI2}
\end{equation}

In order to prove that the FCI wave functions $\Psi'$  and $\Psi''$  are identical (and hence so are their corresponding electronic energies), it must be shown that $W'$ and $W''$ are equal for all $S$  selections. 
	
As a first step, we show that the number of the linearly independent $S$ scenarios is exactly $\binom{m}{n}$ due to the antisymmetrization.
In particular, if any two (or more) $\chi$ functions within of selection $S$ are identical, the corresponding basis product drops out in the course of antisymmetrization, as the same expression is obtained with opposite sign upon variable exchange. 
Furthermore, $S$ selections that only differ in the order of $\chi$ functions necessarily have the same weight up to a sign as the antisymmetrization process interconverts them. 

FCI defined in Eq.~\eqref{eq:FCI}	optimizes $d$ variational parameters explicitly in order to minimize energy.  
According to Eq.~\eqref{eq:W}, the reference-free expansion of FCI of Eq.~\eqref{eq:FCI2} implicitly varies  parameters $W'^{(\prime)}$ of dimension $\binom{m}{n}$   which implies the invariance of reference.

As for truncated CI and CC methods or frozen core approximation, the many-body solution is no longer independent of the reference molecular orbital set. 
The breaking of the invariance is owed to the fact that the dimension of $W'^{(\prime)}$ equals $\binom{m}{n}$ independently  of the applied computational method whereas the dimension of $C'^{(\prime)}$ decreases for truncated approaches, i.e., the summation in Eq.~\eqref{eq:W} runs on a restricted configuration space of dimension $d<\binom{m}{n}$. 
This means that $W'^{(\prime)}$ values become inevitably interdependent, which also prevents them from reaching their optimal values.

\bibliographystyle{achemso}
\bibliography{CC_benchmark}{}
\end{document}